\documentclass[12pt,preprint]{aastex}
\usepackage{epsfig}
\usepackage{bm}

\slugcomment{KSUPT-04/2 \hspace{0.5truecm} March 2004}

\shorttitle{Alfven waves and the CMB}
\shortauthors{Chen et al.}

\begin{document}

\title{Looking for Cosmological Alfv\'en Waves in WMAP Data}

\author{Gang Chen\altaffilmark{1}, Pia Mukherjee\altaffilmark{2}, 
        Tina Kahniashvili\altaffilmark{1,3}, 
        Bharat Ratra\altaffilmark{1}, and Yun Wang\altaffilmark{2}}
\altaffiltext{1}{Department of Physics, Kansas State University, 116 
                 Cardwell Hall, Manhattan, KS 66506.}
\altaffiltext{2}{Department of Physics and Astronomy, University of Oklahoma, 
                 440 W. Brooks Street, Norman, OK 73019.}
\altaffiltext{3}{Center for Plasma Astrophysics, Abastumani Astrophysical 
                 Observatory, A.~Kazbegi ave 2a, Tbilisi, 380060, 
                 Republic of Georgia.}

\begin{abstract}
A primordial cosmological magnetic field induces and supports vorticity 
or Alfv\'en waves, which in turn generate cosmic microwave background (CMB) 
anisotropies. A homogeneous primordial magnetic field with fixed direction 
induces correlations between the $a_{l-1,m}$ and $a_{l+1,m}$ multipole 
coefficients of the CMB temperature anisotropy field. We discuss the 
constraints that can be placed on the strength of such a primordial 
magnetic field using CMB anisotropy data from the WMAP experiment. We place 
3 $\sigma$ upper limits on the strength of the magnetic field of $B < 15$ nG 
for vector perturbation spectral index $n=-5$ and $B<1.7$ nG for $n=-7$.
\end{abstract}


\section{Introduction}

The origin of the large-scale part of observed galactic magnetic fields, of 
$\sim \mu$G (microGauss) strength and apparently coherent over $\sim 10$ kpc
scales, is unknown. They could be the consequence of nonlinear amplification 
of a tiny seed field by galactic dynamo processes. An alternate possibility
is amplification of a weak seed field through anisotropic protogalactic 
collapse and subsequent further amplification via galactic differential 
rotation. In both cases a primordial seed field of strength exceeding 
$10^{-13}$ to $10^{-12}$ G, coherent over $\sim $ Mpc scales, is apparently
needed,  and it is often suggested that upto a $\sim $ nG strength seed 
field might be required. See Kulsrud (1999), Widrow (2002), and Giovannini 
(2003) for reviews of the state of the art in this area. A primordial 
magnetic field of present strength $\sim $ nG can leave observable 
signatures in the cosmic microwave background (CMB) anisotropy.

In standard cosmologies vorticity perturbations decay and so do not 
contribute to CMB temperature or polarization anisotropies. The
presence of a cosmological magnetic field generated during an earlier 
epoch\footnote{
Quantum fluctuations during an early epoch of inflation can generate  
a primordial nG magnetic field, coherent over very large scales (see, 
e.g., Ratra 1992; Bamba \& Yokoyama 2004)}
changes this situation: a primordial magnetic field induces and supports 
vorticity or Alfv\'en waves (Adams et al.~1996; Durrer, Kahniashvili, 
\& Yates 1998, hereafter DKY). These vector perturbations generate CMB 
anisotropies.\footnote{
In the future one can hope to constrain vector modes through their effect
on CMB anisotropy polarization anisotropy spectra. CMB polarization spectra 
that result from vector perturbations due to a primordial magnetic field have 
been discussed by Seshadri \& Subramanian (2001), Pogosian, Vachaspati, \&
Winitzki (2002), Mack, Kahniashvili, \& Kosowsky (2002), and Subramanian,
Seshadri, \& Barrow (2003), while Lewis (2004) considers the case of vector
modes supported by free-streaming neutrinos.} 
The presence of a preferred direction due to a homogeneous magnetic field 
background leads to an $m$ dependence of $\langle a_{lm}a_{lm}^*\rangle$, 
and induces correlations between the $a_{l+1,m}$ and $a_{l-1,m}$ multipole 
coefficients of the CMB temperature anisotropy field. Since the CMB 
anisotropies are observed to be random 
Gaussian\footnote{
Colley, Gott, \& Park (1996), Mukherjee, Hobson, \& Lasenby (2000), and 
Park et al.~(2001) are some early discussions of the Gaussianity of the CMB
anisotropy. More recent discussions of the Gaussianity of the WMAP CMB 
anisotropy data are in Komatsu et al.~(2003), Colley \& Gott (2003), 
Chiang et al.~(2003), Park (2004), Eriksen et al.~(2004a, 2004b), 
Coles et al.~(2004), Vielva et 
al.~(2004), Copi, Huterer, \& Starkman (2003), Hansen et al.~(2004), 
Gurzadyan et al.~(2004), and Mukherjee \& Wang (2004). The simplest inflation 
models predict Gaussian fluctuations (see, e.g., Fischler, Ratra, \&
Susskind 1985; Ratra 1985), and this is consistent with most 
observational indications (see, e.g., Peebles \& Ratra 2003). While there
are indications of mild peculiarities in some subsets of the WMAP data,
for example the apparent paucity of large-scale power (Spergel et al.~2003;
see G\'orski et al.~1998a for a similar indication from COBE data) and the
differences between data from different parts of the sky (see papers cited 
above), foreground contamination (see, e.g., Park, Park, \& Ratra 2002;
Mukherjee et al.~2002, 2003; de Oliveira-Costa et al.~2003; Bennett et 
al.~2003b; Tegmark, de Oliveira-Costa, \& Hamilton 2003) and other 
systematics might be responsible for part of this.}, it is known that such a 
contribution can only be subdominant. 

We use the observed $\overline{\langle a_{l-1,m}a_{l+1,m}^* \rangle}$ 
correlations measured by WMAP to place constraints on the strength of a 
homogeneous primordial magnetic field. The angular brackets here denote
an ensemble average, and the overbar indicates an average over $m$ for 
each $l$. Limited by cosmic variance uncertainties, this would be the useful 
measure to characterize the signature of a homogeneous primordial magnetic 
field.  

The model on which we base our analysis is introduced in \S 2. Our analysis 
of the WMAP data and our results are discussed in \S 3. We conclude in \S 4.

\section{CMB temperature anisotropies generated by Alfv\'en waves}

We assume that the homogeneous magnetic field ${\mathbf B}$ is generated 
prior to the time of recombination. Such a field could be generated during 
the electroweak phase transition (see, e.g., Vachaspati 1991; Sigl, Olinto, 
\& Jedamzik 1997; Giovaninni \& Shaposhnikov 1998), or by an $\alpha$-effect
dynamo driven by collective neutrino-plasma interactions (Semikoz \& 
Sokoloff 2004). The energy density of this field, $B^2/(4\pi)$, must be 
small, to prevent a violation of the
cosmological principle, and so may be treated as a first order perturbation.
Accounting for the high conductivity of the primordial non-relativistic 
plasma (with $v \ll 1$ where ${\bf v}$ is the velocity field of the plasma), 
we may use the infinite conductivity, frozen-in condition, ${\bf E} + {\bf v 
\times  B} = 0$. We also assume that charged particles are tightly coupled 
to the radiation. We write ${\bf B}={\bf B_0} + {\bf B_1}$, where ${\bf B_1}$ 
denotes the first order vector perturbation in the magnetic field 
(where ${\bf \nabla \cdot B_1}=0$), and ${\bf v}= 0 + {\bf \Omega}$ with 
${\bf \Omega}$ being the first order vector perturbation in the fluid 
velocity (where ${\bf \nabla \cdot \Omega}=0$).

As a consequence of magnetic flux conservation the field lines in an 
expanding universe are conformally diluted, $B_0 \propto 1/a^2$, and 
the Alfv\'en velocity in the  photon-baryon plasma during the photon 
dominated epoch when the energy density $\rho_R=\rho_{\gamma} + \rho_b 
\simeq \rho_{\gamma}$ until recombination, is $v_A=B_0/\sqrt{4\pi(\rho_R+p_R)}
= 4 \times 10^{-4} (B_0/10^{-9}\mbox{G})$, and is time independent.  
Rescaling physical quantities according to the expansion of the universe,
the MHD equations result in an equation describing Alfv\'en wave propagation 
with velocity $v_A ({\bf b \cdot {\hat k}}) \equiv v_A \mu$ (DKY)
\begin{equation}
   \ddot{\bf {\Omega}}=v_A^2 ({\bf b \cdot k})^2 \bf {\Omega} .
   \label{alfven-eq}
\end{equation}
Here ${\bf b}\equiv {\bf B_0}/B_0$ is the unit vector in the direction of 
the magnetic field and an overdot denotes a derivative with respect to 
conformal time $\eta$. Choosing only the sine mode, to satisfy the initial 
condition ${\bf\Omega}({\bf k}, \eta=0)=0$, we have
\begin{equation}
   {\bf \Omega}({\bf k}, \eta) \simeq {\bf \Omega_0} v_A k \mu \eta, 
   ~~~~~~~~~|{\bf \Omega}_0| = \frac{v_A}{B_0} |{\bf B}_1| .
  \label{1alfven-sol}
\end{equation}
In eq. (\ref{alfven-eq}) we have neglected viscosity so the vorticity 
solution in eq. (\ref{1alfven-sol}) is applicable only on scales bigger
than the damping scale. Assuming that the initial vector perturbation is
generated by a random process, the two-point correlation function of the 
vorticity field can be written as (Pogosian et al.~2002)
\begin{equation}
   \langle \Omega_{0i}^* ({\bf k}) \Omega_{0j}({\bf k^\prime}) \rangle
   = \left[(\delta_{ij}-{\hat k}_i{\hat k}_j)S(k)
   + i\epsilon_{ijl}{\hat k}_l A(k)\right]
   \delta ({\bf k} - {\bf k^\prime}) .
   \label{two-point}
\end{equation}
Here $\epsilon_{ijl}$ is the totally  antisymmetric tensor, and the $S(k)$ 
and $A(k)$ power spectra describe the symmetric and helical parts of the 
two-point correlation function. We assume that the spectra 
$S(k)$ $(= |{\Omega}_0 (k)|^2)$ and $A(k)$ are given by simple power laws 
of the scale $1/k$ on scales larger than the perturbation damping scale 
$1/k_D$, i.e., for $k < k_D$
we take $S(k) = S_0 {k^n}/{k_D^{n+3}}$ and $A(k) = A_0 {k^m}/{k_D^{m+4}}$.
Here $S_0$ and $A_0$ are dimensionless normalization constants with 
$S_0\geq A_0$, and $n$ and $m$ are spectral indexes. The cutoff scale $1/k_D$ 
is the scale below which the magnetic field is damped away, due to the
non-infinite value of the conductivity. 
Using $T_{\rm dec} \sim 0.3\,{\rm eV}$ 
and $t_{\rm dec} \sim 10^{23}~$cm, the comoving magnetic field damping wave
number at decoupling is $k_D(t=t_{\rm dec}) \sim 3 \times 10^{-10} 
{\mbox{cm}}^{-1}$ (DKY).    

The CMB fractional temperature anisotropy, in direction $\bf n$ on the sky, 
induced by a vorticity perturbation (ignoring a possible dipole contribution 
from the vector perturbation) is (DKY)
\begin{equation}
   \frac{\Delta T}{T}({\bf n}, {\bf k}) \simeq {\bf n} \cdot {\bf \Omega}
   ({\bf k}, \eta) = {\bf n} \cdot {\bf \Omega}_0 v_A \mu (k \eta_{dec}) .
   \label{delta-T}
\end{equation}
Decomposing the CMB fractional temperature anisotropy in a spherical 
harmonic expansion,
\begin{equation}
   \frac{\Delta T}{T}({\bf n})=\sum_{l = 2}^\infty \sum_{m = -l}^l 
   a_{lm}Y_{lm}({\bf n})
\end{equation} 
and using the definition of the power spectrum $C_l$,
\begin{equation} 
  \langle \frac{\Delta T}{T} ({\bf n}) \frac{\Delta T}{T} ({\bf n^\prime})
  \rangle = \frac{1}{4\pi}\sum_{l = 2}^\infty (2l+1)C_lP_l({\bf n} \cdot 
  {\bf n^\prime}),
  \label{power-spectra-def}
\end{equation}
where $P_l$ is the Legendre polynomial, we obtain, in the isotropic case, 
$C_l = \langle a_{lm}^* a_{lm} \rangle$. Here the angular brackets 
denote a theoretical (averaging) expectation value over an ensemble of 
statistically identical universes. In Fourier space this expectation 
value can be replaced by integration over all possible wavenumbers, 
i.e., $\langle ... \rangle \rightarrow \int d^3 {\bf k}/(2\pi)^3$. 

Computing $\langle a_{lm}^* a_{l^\prime m^\prime} \rangle$, it can
be shown that the helical part of vorticity does not contribute (see 
Pogosian et al.~2002). Hence in what follows we consider only the 
symmetric part of the spectrum. Detailed computation of $C_l$'s for
vorticity perturbations are described in DKY where it has been shown 
that the presence of a homogeneous magnetic field induces off-diagonal 
correlations in multipole space, in particular correlations between $l$ 
and $l \pm 2$ multipole coefficients. To quantify this we introduce
a second power spectrum defined by
\begin{equation}
   D_l(m) = \langle a_{l-1,m}^* a_{l+1, m} \rangle =
   \langle a_{l+1,m}^* a_{l-1, m} \rangle . 
   \label{dlm}
\end{equation}
The two power spectra, $C_l(m)$ and $D_l(m)$, depend on the spectral 
index $n$, the normalization constant $S_0$, the Alfv\'en velocity $v_A$ 
and the perturbation damping wavenumber $k_D$. The power spectra are
defined only for $n > -7$ (the quadrupole diverges at small $k$ for 
$n \leq -7$), and for $n > -1$ the results are determined by 
the damping wavenumber $k_D$. The case $n = -5$ corresponds to 
the Harrison-Peebles-Yu-Zel'dovich scale-invariant spectrum ($C_l, D_l 
\sim l^2$). See DKY for a more detailed discussion.

The non-zero correlation of temperature for unequal $l$'s has a simple
physical explanation: The presence of a preferred spatial direction, that
of the magnetic field ${\bf B}_0$, breaks the spatial isotropy of the CMB 
map, leading not only to an $m$ dependence of the correlators, but also 
non-zero off-diagonal (in $l$ space) correlations.\footnote{
In a recent paper, Bershadskii \& Sreenivasan (2004) show that collisions 
between Alfv\'en wave packets and their cascades could generate 
arcminute-scale CMB temperature anisotropies, and argue that this is 
consistent with the WMAP data.} 
The temperature perturbation correlation between two points on the sky 
depends not only on the angular separation between the two points, but also 
on their orientation with respect to the magnetic field. 

A simple observational test to detect (or constrain) the presence of a 
homogeneous magnetic field in the Universe is based on computing the $D_l$ 
spectrum of CMB anisotropy data.\footnote
{Other statistics, related to the $D_l$'s here, could also provide 
useful tests (e.g., Hajian \& Souradeep 2003).}
For this it is useful to introduce the
arithmetic mean over $m$ of the two power spectra,
\begin{eqnarray}
   \overline{C}_l &\equiv &\overline{\langle a_{lm}^* a_{lm} \rangle}=
   \frac{1}{2l+1}\sum_{m=-l}^l \langle a_{lm}^* a_{lm} \rangle
   \nonumber \\
   \overline{D}_l &\equiv &\overline{\langle a_{l-1,m}^* a_{l+1,m} 
   \rangle}= \frac{1}{2l+1}\sum_{m=-l}^l \langle a_{l-1,m}^* a_{l+1,m}
   \rangle  .
   \label{cl-dl-mean}
\end{eqnarray} 
According to DKY, 
\begin{eqnarray}
   {\overline C}_l &\simeq& S_0\left({\eta_{\rm dec}\over
   \eta_0}\right)^2(k_D \eta_0)^{-(n+3)}v_A^2{2^{n+1}\Gamma(-n-1)\over
   3\Gamma(-n/2)^2}l^{n+3}  , ~~~n<-1 \label{Cbar}\\ 
   {\overline C}_l/{\overline D}_l&=&|n+1|
   \left[\frac{\Gamma\left(-\frac{n+1}{2}\right)}
   {\Gamma\left(-\frac{n}{2}\right)}\right]^2 ~~~~
   n<-1.
   \label{Char++}\\
   \overline{C}_l &\simeq& \overline{D}_l \simeq S_0\left({\eta_{\rm dec}\over
   \eta_0}\right)^2(k_D \eta_0)^{-2}v_A^2{1\over n+1}l^2 , 
   ~~~n>-1 \label{Cbar+}
\end{eqnarray}

Using ${\bf B}_1\leq {\bf B}_0$, we have $|{\bf \Omega}_0|^2 k^3 \leq v_A^2$
(see eq.~[2]). This inequality must hold  on all scales inside the Hubble 
radius at decoupling, $k\ge 1/\eta_{\rm dec}$. With the $S(k)$ spectrum 
definition in eq.~(3) we therefore get $2S_0(k/k_D)^{n+3}\leq v_A^2$ for  
$1/\eta_{\rm dec} \le k \le k_D$, implying
\begin{eqnarray}
  2S_0(k_D \eta_{\rm dec})^{-(n+3)} &\leq& v_A^2 ~~~~~ n\le -3
  \label{lim1},\\
  2S_0 &\leq& v_A^2 ~~~~~ n\ge -3 ~.\label{lim2}
\end{eqnarray}
So for $n\leq -3$ the result is independent of the damping wavenumber 
$k_D$. 
 
We now estimate an upper limiting value of $l$, $l_C$, beyond which 
our approximation is no longer valid. Just like scalar perturbations 
(Peebles 1980), vector perturbations are affected by collisionless damping. 
Adding a photon drag force term on the right hand side of the vorticity 
equation (\ref{alfven-eq}), we can see that there are no 
oscillations in the vector perturbation case, and that damping occurs 
on scales slightly larger than the damping scale for scalar perturbations,
when $k_C\eta_{\rm dec} \sim 10$ (DKY), corresponding to $l_C \sim 500$, 
beyond which our approximation breaks down.

Inserting the limiting values given for $S_0$ in eqs.~(\ref{lim1}) and
(\ref{lim2}) in eq.~(\ref{Cbar}), we find\\
for $n = -5$,
\begin{eqnarray}
   \overline{C}_l = 9.04 \times 10^{-16} l^{-2} \left(\frac{B}{1 {\rm nG}}
   \right)^4 , ~~~~~~~~~ \overline{D}_l = \overline{C}_l/2.26 ,
   \label{sol1}
\end{eqnarray} 
and, for $n = -7$,
\begin{eqnarray}
   \overline{C}_l = 8.61 \times 10^{-10} l^{-4} \left(\frac{B}{1 {\rm nG}}
   \right)^4 , ~~~~~~~~~
   \overline{D}_l = \overline{C}_l/2.17 .
   \label{sol2}
\end{eqnarray} 
In this paper, we use $n=-5$ and $n=-7$ as two illustrative cases.
These two cases are interesting as they correspond to a 
Harrison-Peebles-Yu-Zel'dovich scale-invariant spectrum result for 
the $C_l$'s and $D_l$'s, and to a possible inflation model primordial 
vorticity field perturbation spectrum, respectively. These two cases also 
span the range of constraints that can be placed on $B$ using this 
method, in the range $-3 \geq n \geq -7$, i.e., $n=-5$ gives the weakest 
and $n=-7$ the strongest constraint on $B$. With $k_D \eta_0 \sim 0.4 \times
10^{14}$, from eqs.~(9) and (11) $B$ is not meaningfully constrained for
$n > -3$.

\section{Analysis and results}

We use the foreground cleaned Q, V, and, W band co-added WMAP data (Bennett 
et al.~2003a) to determine the off-diagonal correlations. The data are 
available in the Healpix format (Gorski, Hivon, \& Wandelt 1998b) at 
resolution $N_{\rm side}=512$. For each value of the
magnetic field amplitude $B$ we generate 5000 simulations of the CMB sky, 
and each time apply the $Kp2$ Galactic cut mask prior to computing the 
model $D_l$'s. The expected value of the model $D_l$'s, obtained from the 
mean of these simulations, is then compared with the $D_l$'s obtained 
similarly from the WMAP data, using the $\chi^2$ statistic. Confidence 
levels on the field strength are derived from the resulting likelihood 
function. 
 
Specifically, each simulation is a realization of the CMB with power spectrum 
$C_l$ given by the best fit flat-$\Lambda$ CDM model with power-law primordial 
power spectrum (Spergel et al.~2003), and with $D_l$ the same as that 
predicted by eq.~(10) (or more specifically eqs.~[14] or [15]) for a given 
value of $B$. In other words, we generate $a_{lm}$'s such that they satisfy
\begin{equation}
   \langle a_{lm}^* a_{l^\prime m^\prime} \rangle = 
   \delta_{m,m^\prime}\left[ \delta_{l,l^\prime} C_l + (\delta_{l+1,l^\prime-1} 
   + \delta_{l-1,l^\prime+1}) \overline{D}_{l} \right],
\end{equation} 
instead of 
\begin{equation}
   \langle a_{lm}^* a_{l^\prime m^\prime}\rangle = \delta_{m,m^\prime} 
   \delta_{l,l^\prime} C_l.
\end{equation}

The $a_{lm}$'s are generated upto an $l_{\rm max}$ of 512 (corresponding to 
Healpix resolution $N_{\rm side}=256$; since the WMAP data are expected
to contain useful cosmological information upto such an $l_{\rm max}$). 
These are then convolved with the beam functions of each of the Q, V, and W
radiometer channels (8 in all) to produce 8 maps at Healpix resolution 
$N_{\rm side}=512$ (because the noise maps have this resolution). Independent 
Gaussian noise realizations of rms $\sigma_0/\sqrt{N_{\rm obs}}$ from WMAP are added
to the maps, and the 8 maps are co-added weighted by $N_{\rm obs}/\sigma_0^2$, 
where the effective number of observations $N_{\rm obs}$ varies across the 
sky, and $\sigma_0$ is different for each radiometer channel. This is how 
each simulation is created. Hereafter the same analysis procedure that is 
applied to the data map is applied to each of the simulations. This consists 
of bringing the co-added map down to Healpix resolution $N_{\rm side}=256$ 
(since we mostly use $D_l$'s only upto an $l_{\rm max}$ of 300 in the 
subsequent analysis), and applying the $Kp2$ sky cut prior to computing the 
$D_l$'s.

The whole analysis can be repeated for different values of the spectral 
index $n$ that characterizes the spectrum of the magnetic field perturbations.

Fig.~1 shows the $D_l$'s obtained from the WMAP data (crosses) and the median 
and 68\% confidence range contours from simulations for two illustrative values of $B$. 
The spread of about $10^{-3}$ mK$^2$ in the 68\% confidence contours for 
$l (l+1) D_{l}$ is consistent with what is expected from cosmic variance alone.

To compare the likelihoods for different $B$ values, we use the diagonal
$\chi^2$ statistic
\begin{equation}
   \chi^2 = \sum_{l=2}^{300} \frac{ ({D^W}_l - \overline{{D^S}_l})^2 }
   {\sigma_l^2},
\end{equation}
where $D^W_l$ is the WMAP data value, and $\overline{{D^S}_l}$ the 
average and $\sigma_l$ the standard deviation of the ${D^S}_l$, 
both obtained from 5000 model simulations, for each value of $B$. Note 
that we do not use the full covariance matrix for the $D_l$'s but rather
just the diagonal terms. This is because 5000 simulations are not 
sufficient to produce a reliably converged full covariance matrix for
the $D_l$'s. As also noted by Eriksen et al.~(2004b), for example,
even the above $\chi^2$ test can provide a just comparison between
data and simulations. The likelihood is proportional to $e^{- \chi^2 /2}$. 
We calculate the likelihood for a few different values of $B$.

The likelihood function obtained for the $n=-5$ case is shown in Fig.~2. 
After integration, we get a 3 $\sigma$ confidence upper limit of $B < 15$ 
nG. As we can see from Fig.~2, $B$=0 G, which corresponds to pure Gaussian 
primordial fluctuations, is within the 1 $\sigma$ confidence range from the 
peak of the likelihood function (this 1 $\sigma$ range corresponds to a 
$\delta B$ of 3.9 nG). The jaggedness in the likelihood function is from 
the fluctuation of mean $D_l$'s used in calculating $\chi^2$ (see Fig.~1). 
With 5000 simulations, the fluctuation of mean $D_l$'s should be of the order 
of $\sigma_{l}/\sqrt{5000}$. This results in a fluctuation of 0.1 in the 
$\chi^2$ values, or a 5\% uncertainty in the estimated likelihood. This 
does not much affect the results of our analysis (for example for the 3
$\sigma$ limit on $B$ we look for a $\Delta\chi^2$ of 9, which is not very 
sensitive to a 5\% uncertainty in the estimated likelihoods).

For the $n = -7$ case (see Fig.~3), the 3 $\sigma$ confidence limit is 
$B < 1.7$ nG, and 
again the pure Gaussian primordial fluctuation case with $B=0$ G is not far
from the 1 $\sigma$ confidence range from the peak. In this case 1 $\sigma$ 
corresponds to a $\delta B= 0.4$ nG. More stringent limits are obtained in 
this case as expected (eq. [15] shows that larger $D_l$'s with a stronger 
$l$-dependence are predicted by the model for $n=-7$).

The conditions of homogeneity and unidirectionality of the primordial 
magnetic field may be a better approximation on some scales rather than
others. In each of the above cases for $n$, other ranges in $l$, such as 
$2-100$, $101-200$, $201-300$, or $2-500$, did not indicate anything 
qualitatively different, i.e., $B = 0$ G remains a satisfactory fit.

\section{Conclusions}

We study off-diagonal correlations of the form
$D_l=\overline{\langle a_{l-1,m}a_{l+1,m}^* \rangle}$ in the first
year WMAP CMB anisotropy data. Such correlations can result from a 
homogeneous primordial magnetic field. We do not find significant 
off-diagonal correlations in the data, which appear to be satisfactorily 
fit by a zero primordial magnetic field hypothesis. We place 3 $\sigma$ upper 
limits on the strength of the magnetic field of $B < 15$ nG for spectral
index $n=-5$ and $B<1.7$ nG for $n=-7$. These two cases are interesting
as they correspond to a Harrison-Peebles-Yu-Zel'dovich scale-invariant
spectrum result for the $C_l$'s and $D_l$'s, and to a possible inflation 
model primordial magnetic field perturbation spectrum, respectively. 
These two cases also span the range of constraints that can be placed on 
$B$ using this method. Future CMB anisotropy data should allow for tighter 
constraints on a primordial cosmological magnetic field.

\bigskip

We acknowledge useful discussions with R.~Durrer and A.~Kosowsky. GC, TK, 
and BR acknowledge support from NSF CAREER grant AST-9875031 and DOE 
EPSCoR grant DE-FG02-00ER45824. TK also acknowledges CRDF-GRDF grant 3316.
PM and YW acknowledge support from NSF CAREER grant AST-0094335.

\begin{figure}
\epsscale{0.8}\plotone{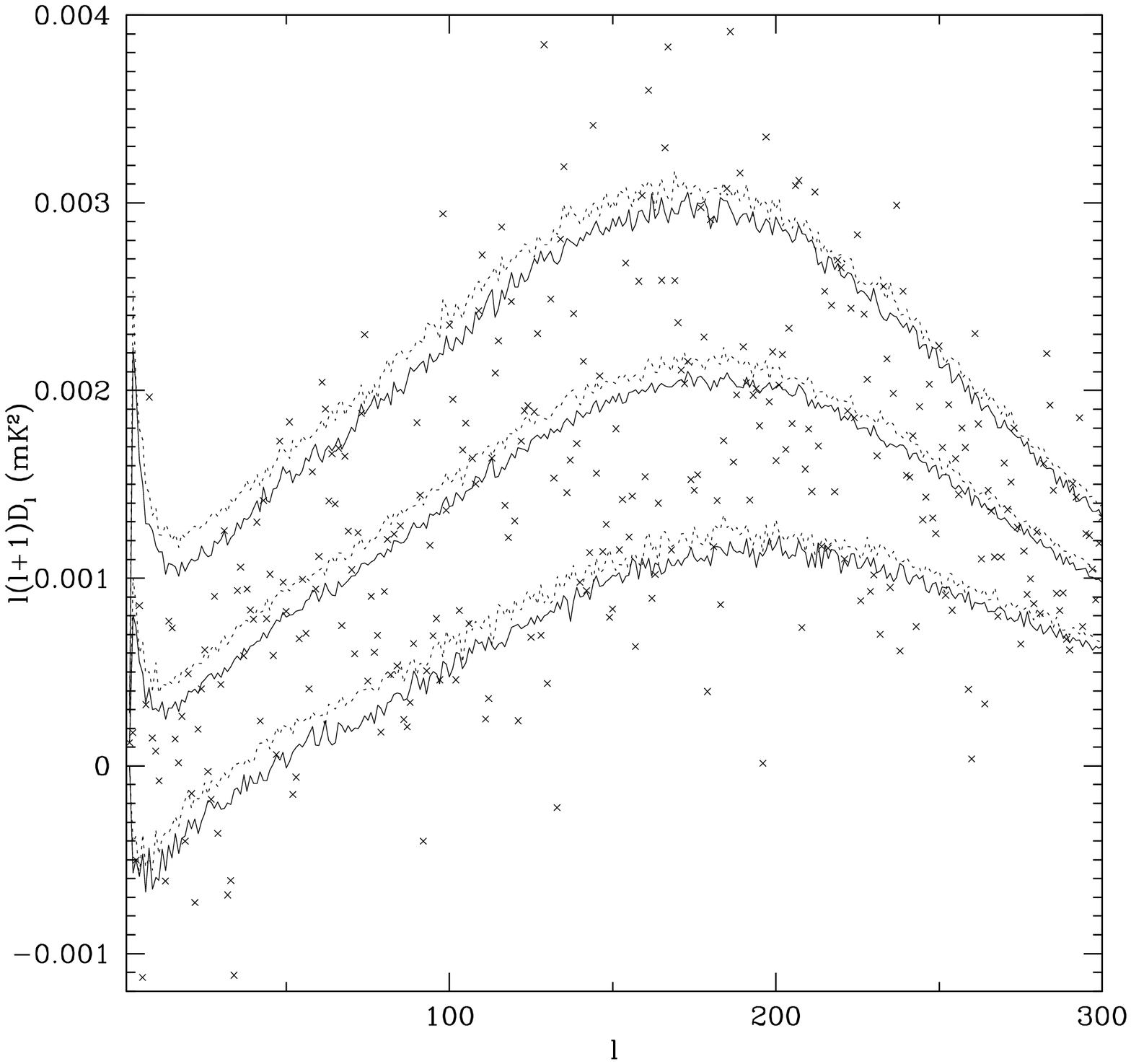}
\figcaption{Off-diagonal power spectra obtained from the WMAP data (crosses) 
and the median and 68\% confidence contours obtained from model simulations 
with magnetic field strengths $B = $ 0 (solid lines) and 16 (dotted lines) 
nG, for a magnetic field perturbation spectral index $n=-5$.}
\end{figure}

\begin{figure}
\epsscale{0.8}\plotone{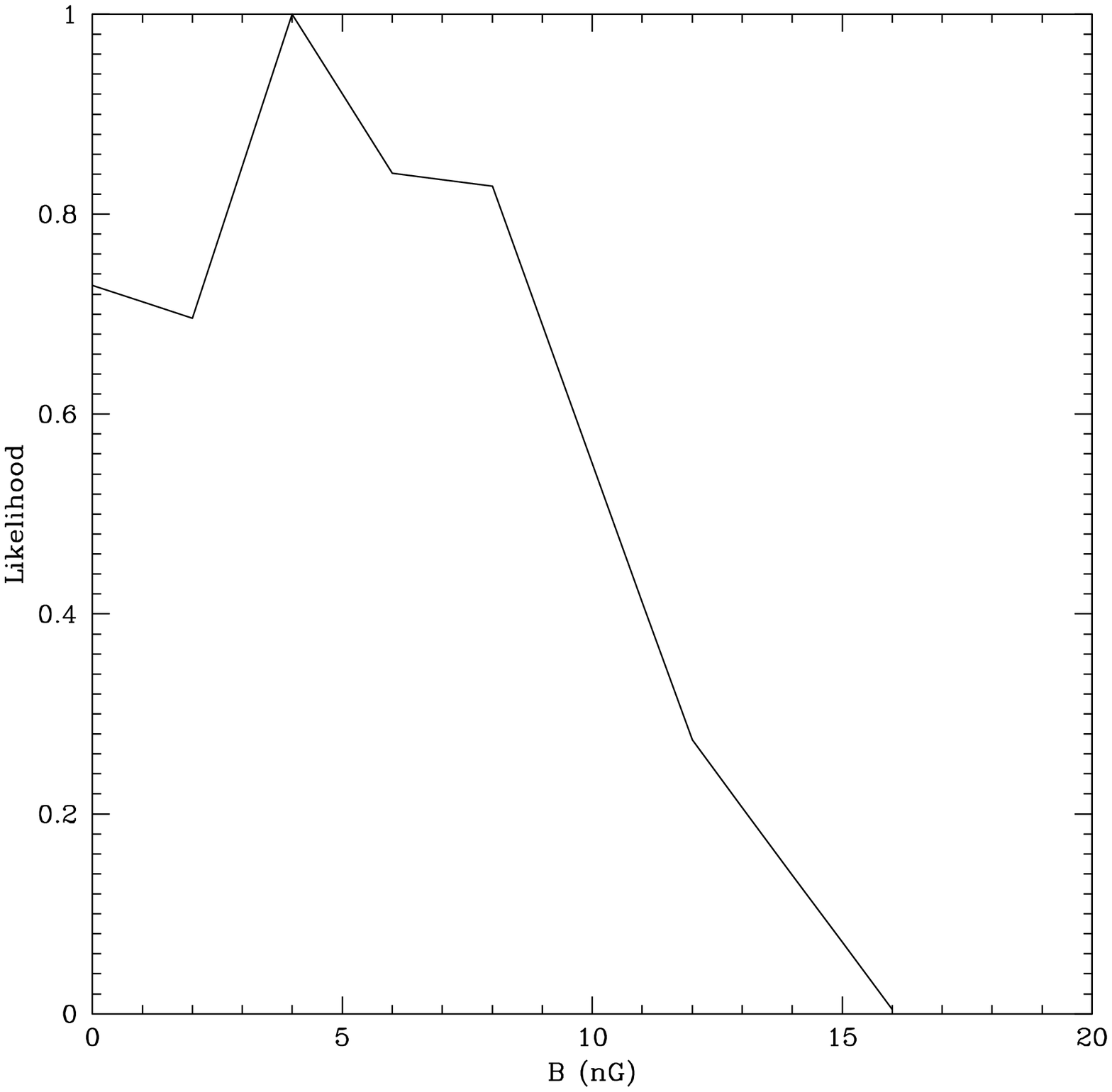}
\figcaption{The likelihood as a function of the strength of the magnetic
field for a magnetic field perturbation spectral index $n=-5$.}
\end{figure}

\begin{figure}
\epsscale{0.8}\plotone{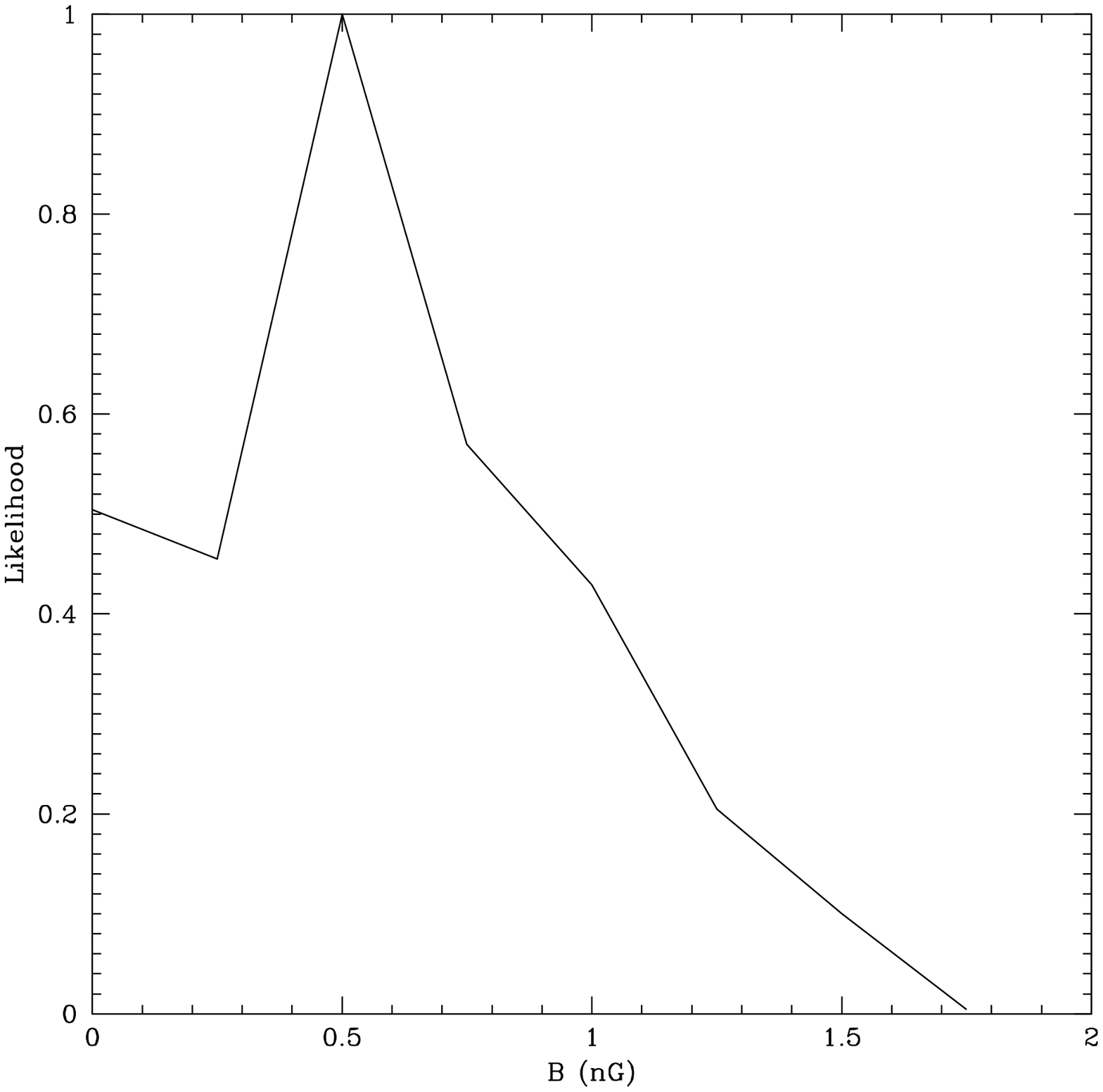}
\figcaption{The likelihood as a function of the strength of the magnetic 
field for a magnetic field perturbation spectral index $n=-7$.}
\end{figure}

\end{document}